\documentclass[a4paper, conference]{IEEEtran}

\usepackage[T1]{fontenc}
\usepackage[utf8]{inputenc}

\usepackage{amsmath}
\usepackage[cmintegrals]{newtxmath}
\usepackage{bm}
\usepackage{cite}
\usepackage{color}
\usepackage{graphicx}
\usepackage{glossaries}
\usepackage{booktabs}
\usepackage{dcolumn}
\usepackage{url}
\usepackage{siunitx}
\usepackage{tablefootnote}
\DeclareSIUnit \belc {Bc}
\DeclareSIUnit{\belmilliwatt}{Bm}
\DeclareSIUnit{\dBm}{\deci\belmilliwatt}
\DeclareSIUnit{\dBc}{\deci\belc}
\DeclareSIUnit{\sample}{Sa}

\usepackage{placeins}

\usepackage{tikz}
\usetikzlibrary{positioning}
\usetikzlibrary{shapes}
\usetikzlibrary{fit}
\usetikzlibrary{spy}
\tikzset{>=latex}
\pgfdeclarelayer{bg}
\pgfdeclarelayer{fg}
\pgfsetlayers{bg,main,fg}

\usepackage{pgfplots}
\pgfplotsset{compat=1.16}

\makeglossaries

\newacronym{FPGA}{FPGA}{field-programmable gate array}
\newacronym{PC}{PC}{personal computer}
\newacronym{SDR}{SDR}{software-defined radio}
\newacronym{Rx}{Rx}{receive}
\newacronym{Tx}{Tx}{transmit}
\newacronym{ADC}{ADC}{analog-to-digital converter}
\newacronym{PEP}{PEP}{peak envelope power}
\newacronym{USRP}{USRP}{universal software radio peripheral}
\newacronym{DUT}{DUT}{device under test}
\newacronym{SNR}{SNR}{signal-to-noise ratio}
\newacronym{LTI}{LTI}{linear time-invariant}
\newacronym{FFT}{FFT}{fast Fourier transform}
\newacronym{UPOLS}{UPOLS}{uniformly partitioned overlap-save}
\newacronym{OLS}{OLS}{overlap-save}
\newacronym{MIMO}{MIMO}{multiple input multiple output}
\newacronym{DPDK}{DPDK}{Data Plane Development Kit}
\newacronym{UHD}{UHD}{USRP Hardware Driver}
\newacronym{NIC}{NIC}{network interface controller}
\newacronym{GPU}{GPU}{graphics processing unit}
\newacronym{CHDR}{CHDR}{condensed hierarchical datagram for RFNoC}
\newacronym{TDL}{TDL}{tapped delay line}
\newacronym{CPU}{CPU}{central processing unit}
\newacronym{RCU}{RCU}{read-copy-update}
\newacronym{IPI}{IPI}{inter-processor interrupt}
\newacronym{LKML}{LKML}{Linux kernel mailing list}
\newacronym{LTE}{LTE}{Long-Term Evolution}
\newacronym{UE}{UE}{user equipment}

\pgfplotsset{
height=6cm,
width=8.5cm,
}

\begin{document}

\title{Low-Latency Analog-to-Analog Signal Processing using PC Hardware and USRPs}

\author{Maximilian~Engelhardt\IEEEauthorrefmark{2}, Carsten~Andrich\IEEEauthorrefmark{1}\IEEEauthorrefmark{2}, Alexander~Ihlow\IEEEauthorrefmark{1}, Sebastian~Giehl\IEEEauthorrefmark{1}, Giovanni~Del~Galdo\IEEEauthorrefmark{1}\IEEEauthorrefmark{2} \\
\\
\IEEEauthorblockA{\IEEEauthorrefmark{1}Technische Universität Ilmenau, Institute for Information Technology, Ilmenau, Germany}
\IEEEauthorblockA{\IEEEauthorrefmark{2}Fraunhofer Institute for Integrated Circuits IIS, Ilmenau, Germany}%
}

\maketitle

\begin{abstract}
In this paper, we implement a low-latency rapid-prototyping platform for signal processing based on software-defined radios (SDRs) and off-the-shelf PC hardware.
This platform allows to evaluate a wide variety of algorithms in real-time environments, supporting new developments in the fields of classical, AI-based, and hybrid signal processing.
To accomplish this, the streaming protocol of the used USRP X310 devices is implemented using the Data Plane Development Kit (DPDK), which allows to handle network communication in userspace only.
This bypasses the kernel and thus avoids the latencies caused by interrupt handling, scheduling, and context switches.
It allows signal processing to be performed on isolated processor cores that are protected from interrupts to a great extent.
To validate our approach, linear time-invariant channel emulation has been implemented.
For this, an analog-to-analog latency of \SI{31}{\micro\second} was achieved, demonstrating that our PC-based approach enables the implementation of rapid-prototyping systems with low latency.

\end{abstract}


\section{Introduction}

Signal processing applications requiring low latency and deterministic timing are typically realized using \glspl{FPGA}, involving significant implementation effort.
\Glspl{SDR} with standard drivers are focused on the easy implementation of algorithms, but make it difficult to meet real-time requirements.
Here, we present a platform based on PC hardware and the widely used \gls{SDR} \textit{USRP X310}, manufactured by \textit{Ettus Research}, with a software implementation tuned for low latency.

For this, communication with the SDR is not handled via the standard driver, but an optimized implementation of the protocol is used that bypasses the kernel network stack.
Furthermore, the operating system is configured to interfere as little as possible with the application running on isolated CPU cores.
As a test case, LTI channel emulation is implemented, although the platform allows to evaluate a wide variety of algorithms, supporting new developments in the field of AI-based and hybrid signal processing.

\subsection{State of the Art}

For the used \glspl{SDR}, a driver package is provided by the manufacturer, the so-called \gls{UHD}.
To enable low-latency data processing, it also supports the use of the \gls{DPDK}, a framework for optimized network communication~\cite{UHD_DPDK}.
The performance of this interface is evaluated in~\cite{UHD_DPDK_Analysis}:
A \gls{LTE} \gls{UE} is implemented, whereby a data rate of \SI{30}{\mega\bit\per\second} is achieved.
Rather than the overall latency of the system, the jitter of the packet processing is analyzed.
Employing the DPDK allows to reduce it to as little as \SI{179}{\micro\second}, while the best result achieved without DPDK was \SI{448}{\micro\second}.

As an example of low-latency signal processing implemented directly on an FPGA,~\cite{Schwind} can be considered, where a channel emulation is implemented directly on the FPGA of the SDR.
This allows for an outstanding performance, a latency of only \SI{3.5}{\micro\second} at \SI{100}{\mega\sample\per\second}.
However, this approach is limited to the resources offered by a single device:
In this case, the available ports restrict it to 2x2 \gls{MIMO} and the implemented signal processing algorithm is fixed (\gls{TDL} with 18 taps).

\section{Software Implementation}

\subsection{UHD}

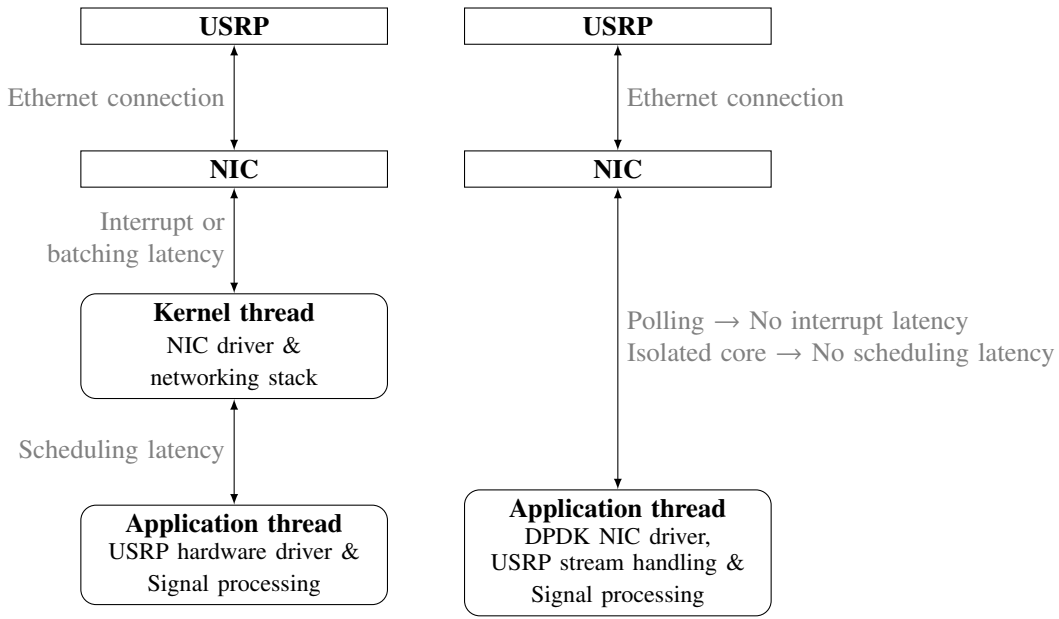
\begin{figure*}[ht!]
\begin{center}
\begin{tikzpicture}[node distance=1.4cm,
component/.style={draw, text width=3.8cm, align=center},
thread/.style={draw, rounded corners=0.2cm, text width=3.8cm, align=center},
arrownodea/.style={midway, left, gray },
arrownodeb/.style={midway, right, gray },
]

\node[component] (usrpa) {\textbf{USRP}};
\node[component, below=of usrpa] (nic) {\textbf{NIC}};
\node[thread, below=of nic] (ktread) {\textbf{Kernel thread} \\ \small{NIC driver \& networking stack}};
\node[thread, below=of ktread] (atread) {\textbf{Application thread} \\ \small{USRP hardware driver \&  \\ Signal processing}};

\draw[<->] (usrpa) -- (nic) node[arrownodea] {Ethernet connection};
\draw[<->] (nic) -- (ktread) node[arrownodea, text width=4cm, align=right] {Interrupt or \\ batching latency};
\draw[<->] (ktread) -- (atread)  node[arrownodea] {Scheduling latency};

\node[component, right=1cm of usrpa] (usrpb) {\textbf{USRP}};
\node[component, below=of usrpb] (nicb) {\textbf{NIC}};
\node[thread, below=4cm of nicb] (treadb) {\textbf{Application thread} \\ \small{DPDK NIC driver, \\ USRP stream handling \& \\ Signal processing}};

\draw[<->] (usrpb) -- (nicb) node[arrownodeb] {Ethernet connection};
\draw[<->] (nicb) -- (treadb) node[arrownodeb, text width=6cm] {Polling $\rightarrow$ No interrupt latency \\ Isolated core $\rightarrow$ No scheduling latency};

\end{tikzpicture}
\end{center}
\caption{On the left is the unoptimized communication path when using the UHD, whereby the \gls{NIC} reports incoming packets via an interrupt to the kernel, which then forwards them to the userspace.
On the right is the optimized communication with DPDK, in which the application polls incoming packets directly from the \gls{NIC}, eliminating all context switches.
}
\label{fig:PacketPath}
\end{figure*}

The UHD is a common driver for the broad portfolio of \gls{SDR} devices manufactured by \textit{Ettus Research}, which are connected via various interfaces.
Since the model used for our experiments, X310, is connected via Ethernet, it is discussed here specifically.
The network connection is established via the conventional kernel network stack.
However, instead of the socket API, an interface based on \texttt{boost::asio} is used to avoid copies of the transferred data.

This method is not optimal for low-latency signal processing, since delays may occur on several levels (see Figure~\ref{fig:PacketPath}):
First, the kernel must be notified about incoming packets which is done via an interrupt, introducing a delay.
In high-load scenarios, interrupt coalescing may be used (see~\cite{InterruptMitigation}), whereby packets are processed in batches, resulting in an even higher latency, which is why this optimization should be deactivated in real-time applications.
Then, the packet is processed in the kernel and passed to the waiting application thread.
In this step, a further delay arises from the necessary scheduling and second context switch.

Analogously, recurring context switches between application software and kernel are also required when sending samples.
In addition, the user does not have full control over the host-side Tx buffers managed by the UHD~\cite{UHD_DPDK_Analysis}.
This is problematic because buffered samples increase the latency of the system.
Therefore, for a low-latency application, Tx-side buffering on the host should be avoided.

\subsection{UHD DPDK}

One possible solution to the problem described above is to use the DPDK, which the UHD supports as an alternative network interface.
This software framework allows network communication to be implemented entirely in userspace, bypassing the kernel network stack.
To achieve this, it provides its own userspace-only drivers for a wide range of NIC types.
These are based on polling so that the delay associated with interrupts can be circumvented.
A DPDK application is executed on isolated CPU cores, which it fully occupies and does not share with other threads.
This completely avoids delays due to scheduling and context switching.
In addition, the DPDK offers other functionalities that support the development of optimized applications, such as efficient memory management and queues.

DPDK integration is implemented in the UHD in such a way that each device is assigned a core that handles only the network communication.
The samples are then exchanged via queues with the application threads running on other cores, where the actual signal processing takes place.

The accompanying documentation provides instructions for moving system interrupts away from the used cores~\cite{UHD_DPDK_Doc}.
Additionally, the performance of the used \gls{CPU} cores was increased by deactivating the respective hyper-threading siblings.

\subsection{Pure DPDK Protocol Implementation}

To circumvent the problems described above, we implemented the sample streaming protocol directly in the DPDK-based application.
This bypasses the UHD in the performance-critical path and allows integrating the reception and transmission of packets into the signal processing flow, eliminating the need for separate IO threads.
To achieve this, it is necessary to implement the sample streaming protocol of the USRP, which is outlined in the following.

\subsection{The CHDR Protocol} \label{subsec:protocol}
The USRP devices rely on the proprietary protocol designated as \gls{CHDR} described in~\cite{RFNoCSpec}.
Since this documentation does not cover all relevant aspects, it was necessary to rely on reverse-engineering of the open-source UHD in some cases.
Thereby, only the actual sample streaming was re-implemented:
The highly complex initialization of the device was done with the UHD for our experiments, as this is not performance relevant.

\begin{figure}[ht!]
\begin{center}
\scalebox{0.35}{\includegraphics{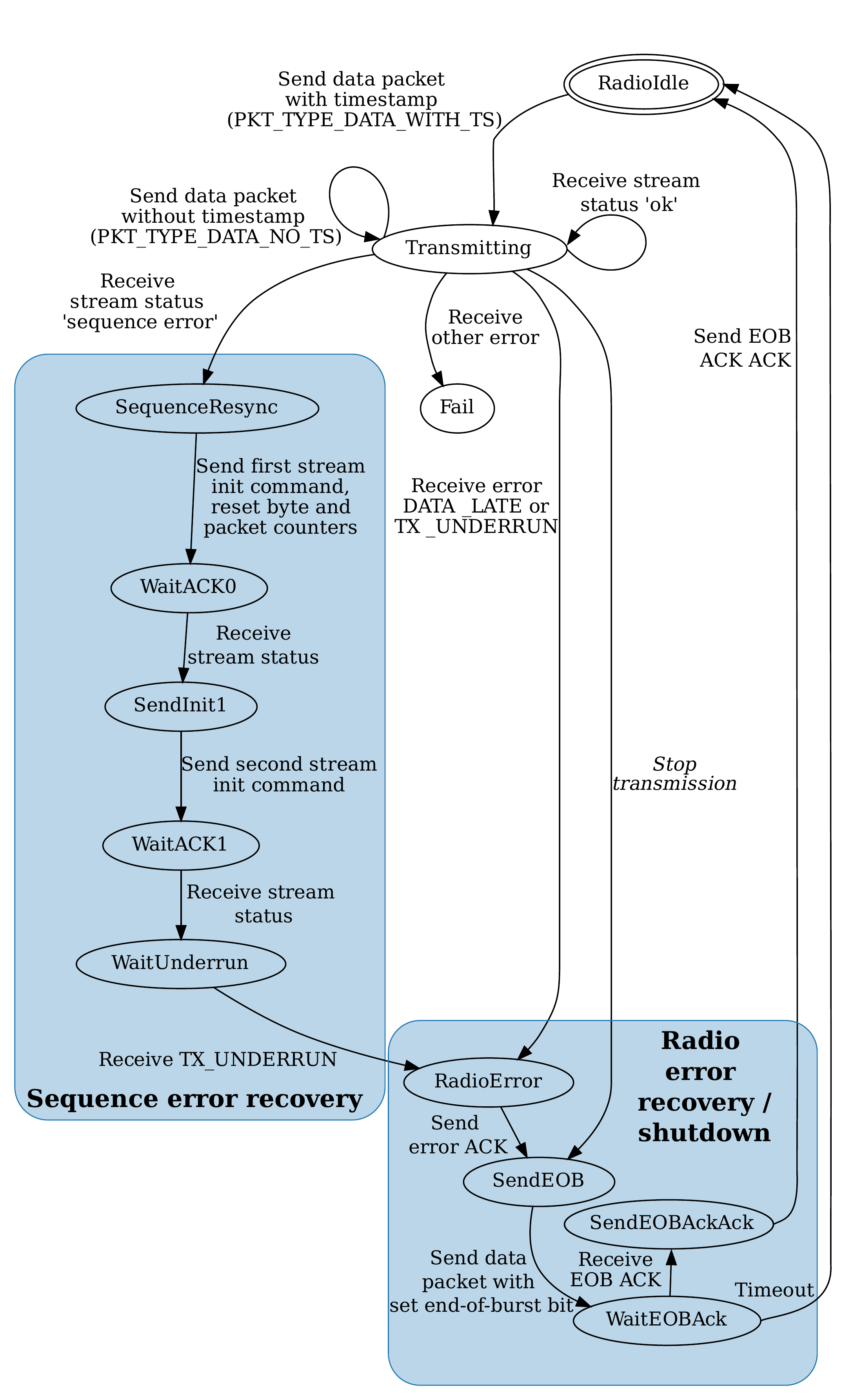}}
\end{center}
\caption{
  Host-side state machine of a Tx stream in the CHDR protocol.
  A lost or delayed sample packet leads to the "sequence error" state, which requires 3 round-trips to fix.
}
\label{fig:StateMachine}
\end{figure}

The analysis of the USRP's protocol has shown that is not designed for real-time applications:
As shown in Figure~\ref{fig:StateMachine}, in case of a Tx sample packet arriving too late, a complex recovery process has to be triggered to continue sample-synchronous streaming, requiring 3 round-trips.
This cannot be solved by adding a timestamp to each sample packet:
The device will still discard any further arriving packets until the error-handling process is completed.

\subsection{Latency Optimization}
To achieve the lowest possible latency, not only the application itself but also the environment must be optimized. 
It is of great importance that no other tasks are running on the \gls{CPU} cores that perform the real-time signal processing.

The basic measure here is to set the scheduling policy \texttt{SCHED\_FIFO}, which is optimized for real-time applications, and to ensure with the boot parameter \texttt{isolcpus} that no other processes are scheduled on these cores.

The next steps are to avoid interference not only from other processes but from the kernel itself.
There are several options that can be activated via boot parameters:

\begin{itemize}
  \item The periodic scheduling clock can be disabled with \texttt{nohz\_full}~\cite{NOHZDoc}.
  \item \texttt{irqaffinity} allows setting which cores handle hardware interrupts~\cite{LinuxParams}. The real-time cores should be excluded here.
  \item The kernel uses \gls{RCU} for lock-less mutual exclusion. This requires to schedule callbacks for later execution. With the options \texttt{rcu\_nocbs} and \texttt{rcu\_nocb\_poll}, the behavior can be optimized to move this task away from the real-time cores~\cite{LinuxRCU}.
\end{itemize}

However, all these measures only mitigate interrupts but do not prevent them completely.
The recent activity on the \gls{LKML} shows that there are efforts by some developers to create a complete solution here~\cite{IsolLKML}, but no inclusion in the mainline kernel seems foreseeable yet.

\section{Evaluation Method}

\subsection{Hardware}

\begin{figure}[ht!]
\begin{center}
\scalebox{0.8}{
\begin{tikzpicture}[node distance=1.9cm,
component/.style={draw, text width=2.5cm, align=center},
thread/.style={draw, rounded corners=0.2cm, text width=3.8cm, align=center},
arrownodea/.style={midway, left, gray },
arrownodeb/.style={midway, right, gray },
]

\node[component] (usrpa) {\textbf{USRP 1} \\ X310 \\ 2x UBX DB};

\node[gray, component, below=of usrpa] (usrpb) {\textbf{USRP 2} \\ X310 \\ 2x UBX DB};

\node[component, right=of usrpa.north east, anchor=north west, minimum height=5cm, text width=2cm] (host) {\textbf{Host PC}};

\draw[gray, ->] ([xshift=0.4cm]usrpa.south) -- ([xshift=0.4cm]usrpb.north) node[right, midway] {1 PPS};
\draw[gray, ->] ([xshift=-0.4cm]usrpa.south) -- ([xshift=-0.4cm]usrpb.north) node[left, midway] {10 MHz};

\draw[<->] (usrpa) -- (usrpa-|host.west) node [above, midway] {Ethernet};
\draw[gray, <->] (usrpb) -- (usrpb-|host.west) node [above, midway] {Ethernet};

\draw[<-] ([yshift=0.2cm]usrpa.west) -- ++(-1cm,0cm) node [midway, above] {Rx1};
\draw[->] ([yshift=-0.2cm]usrpa.west) -- ++(-1cm,0cm) node [midway, below] {Tx1};

\draw[gray, <-] ([yshift=0.2cm]usrpb.west) -- ++(-1cm,0cm) node [midway, above] {Rx2};
\draw[gray, ->] ([yshift=-0.2cm]usrpb.west) -- ++(-1cm,0cm) node [midway, below] {Tx2};

\end{tikzpicture}}
\end{center}
\caption{Test setup consisting of the host PC and two SDRs.
The second SDR, drawn in gray, will only be used for MIMO tests.
}
\label{fig:TestSetup}
\end{figure}
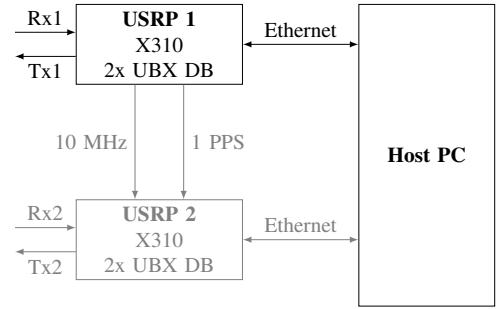

As shown in Figure~\ref{fig:TestSetup}, the core of our test setup is a host PC fitted with a CPU of the type \textit{AMD Ryzen 9}, which offers 16 physical cores.
Two SDRs of type \textit{Ettus Research USRP X310} are connected to it, whereby each is fitted with two daughterboards of type \textit{UBX}.
Only one channel is used per daughterboard to avoid crosstalk between the channels.
For single-channel tests, only one of the \glspl{SDR} is used, while both devices together allow experiments with 2x2 \gls{MIMO}.
The connection to the host is established via one 10 GBit Ethernet link each.
The two SDRs are synchronized by connecting their 1 PPS and \SI{10}{\mega\hertz} ports.

As measurement device, we use a broadband channel sounding system which is also based on the USRP X310.
This allows for measuring the analog-to-analog latency of the overall system, which is the central parameter for our evaluation.
Furthermore, it allows evaluating the correctness of the channel emulation:
If the implementation is correct, the measured channel corresponds to the emulated one.

We use a sample rate of \SI{100}{\mega\sample\per\second} and a center frequency of \SI{3.75}{\giga\hertz} for our experiments.
Another important parameter is the number of samples per packet:
In the default setting, the UHD selects the highest possible value here, which minimizes the overhead, but increases latency.
For our measurements, we set 512 samples per packet, which results in a packetization delay of \SI{5.12}{\micro\second}.

\subsection{Pass-Through}
The most basic test case is the unchanged transmission of the received signal. 
This implies that no signal processing takes place in the host, but the application only forwards the received samples.
This elementary example allows determining the minimum achievable analog-to-analog latency for this system architecture.

\subsection{LTI Channel Emulation}
As a second test case, the emulation of a \gls{LTI} channel is implemented.
Mathematically, this means that the received signal is convolved with a given channel impulse response $h(t)$.
According to the convolution theorem, the convolution in time domain corresponds to the point-wise product in frequency domain.
From this insight, the class of so-called fast convolution algorithms was developed, which is described and evaluated in~\cite{Filter}.
Here, \gls{UPOLS} is to be implemented, which offers high performance, both in throughput and latency, with modest implementation effort.
It is based on the well-known \gls{OLS} method, dividing both the signal and the channel impulse response into segments of equal length.
This segmentation matches the behavior of the \gls{SDR} hardware, which sends and receives samples in blocks (packets) of configurable length.

Another advantage of the method is that it can be used very efficiently for \gls{MIMO} channel emulation:
The number of required \glspl{FFT} scales with the number of ports and not with the number of channels, which is much higher in a fully meshed system.
An additional advantage is that the channel impulse response can be replaced on-the-fly, which allows the emulation of slowly time-variant channels.

In our case, a segment length of 256 samples is used, which means that the hardware is configured to work with packets of this size.
The emulated channel is 512 samples long, which corresponds to \SI{5.12}{\micro\second} at the sample rate of \SI{100}{\mega\sample\per\second} used here.

\section{Results}
For evaluating the performance of the presented approach, achievable analog-to-analog latency is the central measure.
Furthermore, the distribution of the processing delays of individual packets is to be examined to identify further optimization potentials.

\subsection{Analog-to-analog Latency}
The measured analog-to-analog latencies for the pass-through test case are given in Table~\ref{tab:Results}.
The pure DPDK implementation achieved an analog-to-analog latency of \SI{25.2}{\micro\second}, less than a quarter of the result of the UHD-based approach.
This significant performance improvement shows the high potential of this method.
In contrast, DPDK-UHD failed to meet the authors' expectations:
Here, stable transmissions could not be established at the desired sample rate of \SI{100}{\mega\sample\per\second}, so measurements had to be performed at a reduced sample rate of \SI{50}{\mega\sample\per\second} (doubling the packetization delay to \SI{10.24}{\micro\second}).
The latency achieved was only marginally better than that achieved with UHD without DPDK.

\begin{table}[ht!]
\label{tab:Results}
\begin{center}
\caption{Measurement results}

\begin{tabular}[h]{cS[table-format=3.1]S[table-format=3.1]}
 & {Analog-to-analog Latency} & {Sample Rate} \\
UHD & \SI{104.8}{\micro\second} & \SI{100}{\mega\sample\per\second}  \\
UHD-DPDK & \SI{94.8}{\micro\second} & \SI{50}{\mega\sample\per\second} \\
Pure DPDK & \SI{25.2}{\micro\second} & \SI{100}{\mega\sample\per\second}  \\
\end{tabular}

\end{center}
\end{table}

For the LTI channel emulation test case, a latency of \SI{31}{\micro\second} was achieved.
This shows that very good performance can still be achieved when the sample data is actually processed on the CPU instead of passing it on untouched.

\subsection{Packet Processing Delay}\label{sec:ProcessingDelay}

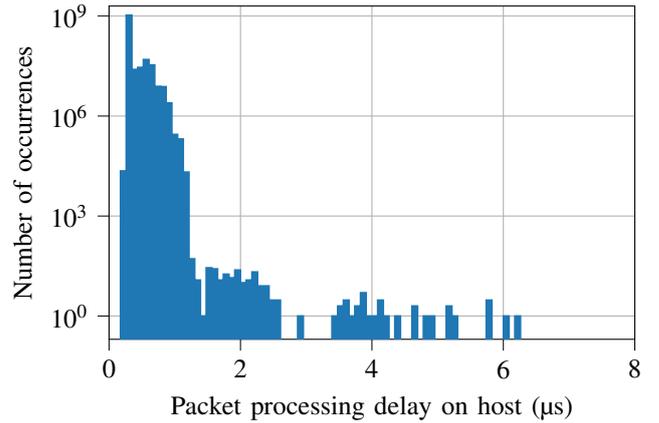
\begin{figure}[ht!]
\begin{center}
\begin{tikzpicture}

\definecolor{color0}{rgb}{0.12156862745098,0.466666666666667,0.705882352941177}

\begin{axis}[
grid style={white!69.01960784313725!black},
ymode=log,
log basis y={10},
tick align=outside,
tick pos=left,
x grid style={white!69.01960784313725!black},
xlabel={Packet processing delay on host (µs)},
xmin=0, xmax=8,
xtick style={color=black},
y grid style={white!69.01960784313725!black},
ylabel={Number of occurrences},
ymin=0.2,
ymax=2e9,
ytick style={color=black},
xmajorgrids, ymajorgrids,
]
\addplot [semithick, color0, const plot mark right, fill]
table {%
0 1e-20
0.0869565217391304 1e-20
0.173913043478261 1e-20
0.260869565217391 22600
0.347826086956522 1061759465
0.434782608695652 25143817
0.521739130434783 28878062
0.608695652173913 49685876
0.695652173913043 34279686
0.782608695652174 7779075
0.869565217391304 7629519
0.956521739130435 2478665
1.04347826086957 281114
1.1304347826087 205188
1.21739130434783 20831
1.30434782608696 52
1.39130434782609 12
1.47826086956522 1
1.56521739130435 28
1.65217391304348 26
1.73913043478261 12
1.82608695652174 18
1.91304347826087 14
2 24
2.08695652173913 10
2.17391304347826 12
2.26086956521739 21
2.34782608695652 8
2.43478260869565 8
2.52173913043478 3
2.60869565217391 3
2.69565217391304 1e-20
2.78260869565217 1e-20
2.8695652173913 1e-20
2.95652173913043 1
3.04347826086956 1e-20
3.1304347826087 1e-20
3.21739130434783 1e-20
3.30434782608696 1e-20
3.39130434782609 1e-20
3.47826086956522 1
3.56521739130435 2
3.65217391304348 3
3.73913043478261 1
3.82608695652174 2
3.91304347826087 5
4 1
4.08695652173913 1
4.17391304347826 3
4.26086956521739 1
4.34782608695652 1e-20
4.43478260869565 1
4.52173913043478 1e-20
4.60869565217391 1e-20
4.69565217391304 2
4.78260869565217 1e-20
4.8695652173913 1
4.95652173913043 1
5.04347826086957 1e-20
5.1304347826087 1e-20
5.21739130434783 2
5.30434782608696 1
5.39130434782609 1e-20
5.47826086956522 1e-20
5.56521739130435 1e-20
5.65217391304348 1e-20
5.73913043478261 1e-20
5.82608695652174 3
5.91304347826087 1e-20
6 1e-20
6.08695652173913 1
6.17391304347826 1e-20
6.26086956521739 1
6.34782608695652 1e-20

};

\end{axis}

\end{tikzpicture}
\end{center}
\caption{Histogram of measured per packet \gls{Rx} to \gls{Tx} delay in the LTI channel emulation test case. Remaining interrupts, while being rare, vastly increase the worst-case delay, limiting the achievable analog-to-analog latency. }
\label{fig:DelayHistogramm}
\end{figure}

Figure~\ref{fig:DelayHistogramm} depicts a histogram of the delay from the reception of an input sample packet until the associated output sample packet is sent in the DPDK application.
It shows that the vast majority of packets were processed in under \SI{1}{\micro\second}, but very rare events caused delays of up to \SI{6.2}{\micro\second}.
This is caused by remaining interrupts, which cannot be completely suppressed with the isolation techniques presented.

\section{Conclusion}
In this paper, we investigated the possibility of implementing a low-latency signal processing system based on USRPs and off-the-shelf PC hardware.
In doing so, the hardware driver provided by the manufacturer UHD was replaced by an optimized custom solution based on the DPDK framework.
With this, a very good performance was achieved, reducing the pass-through latency from \SI{104.8}{\micro\second} to only \SI{25.2}{\micro\second}.
The performance of the UHD's own DPDK interface could not match this.
The resulting software shows the potential of our approach but has to be seen as a prototype so far.
The employed hardware is capable enough for MIMO signal processing.
Overall, it was shown that a rapid-prototyping platform for low-latency signal processing can be implemented with the combination of PC hardware and SDRs.

\section{Future Work}

As was discussed in Section~\ref{sec:ProcessingDelay}, the presented isolation techniques are not sufficient to completely isolate the real-time application from kernel interference.
Since for a continuous output stream every sample packet must arrive in time, the achievable system latency is limited by the worst-case delay.
This is particularly important for more complex applications:
The longer the computing time, the higher the probability that the processing of a single packet can be disturbed by several interrupts.
Therefore, solving this incomplete isolation is an important starting point for further optimization.

Furthermore, it must be noted that the \gls{CHDR} protocol is not designed for coherent sampling in real-time applications as described in Section~\ref{subsec:protocol}.
A modified protocol design, whereby the device merely replaces missing packets in the \gls{Tx} stream with zeros, would significantly shorten the resulting gap in the output signal.
Also, a desirable protocol feature would be to allow distributing \gls{Rx} samples to multiple hosts, for example via the Ethernet multicast mechanism.
This is not provided by the UHD's protocol, obstructing the scaling of the system to multiple hosts.
Overall, an SDR optimized for low-latency real-time applications should use a protocol adapted to it.

Another possible further development would be the integration of the GPU, where DPDK enables a low-latency zero-copy interface.
This would involve additional effort but would provide the enormous computational power of the GPU, which is an invaluable enhancement for AI-based algorithms.


\bibliographystyle{IEEEtran}
\bibliography{paper}

\end{document}